\documentstyle[aps,prl,multicol]{revtex}
\input {epsf.sty}
\begin{document}
\draft
\title{Exclusion Statistics in a trapped two-dimensional Bose gas}
\author{T. H. Hansson}
\address{Department of Physics, University of
Stockholm, Box 6730,\\ S-11385 Stockholm, Sweden }
\author{J. M. Leinaas}
\address{Department of Physics,  University of Oslo,
P.O. Box 1048 Blindern, N-0316 Oslo, Norway }
\author{S. Viefers}
\address{Nordita, Blegdamsvej 17, DK-2100 Copenhagen, Denmark}
\date{\today}
\maketitle
\begin{abstract}
We study the statistical mechanics of a two-dimensional gas with a 
repulsive delta function
interaction, using a mean field approximation. By a direct counting 
of states we establish
that this model obeys exclusion statistics and is equivalent to an 
ideal exclusion statistics
gas.
\end{abstract}
\vspace{-11pt}
\begin{multicols}{2}

\newcommand{\half}{\frac 1 2 }
\newcommand{\eg}{{\em e.g.} }
\newcommand{\ie}{{\em i.e.} }
\newcommand{\etc} {{\em etc.}}

\newcommand{\noi}{\noindent}
\newcommand{\etal}{{\em et al.\ }}
\newcommand{\cf}{{\em cf. }}

\newcommand{\dd}[2]{{\rmd{#1}\over\rmd{#2}}}
\newcommand{\pdd}[2]{{\partial{#1}\over\partial{#2}}}
\newcommand{\pa}[1]{\partial_{#1}}
\newcommand{\pref}[1]{(\ref{#1})}

\newcommand{\bea}{\begin{eqnarray}} 
\newcommand{\eea}{\end{eqnarray}}
\newcommand{\e}{\varepsilon} 
\newcommand{\D}{\partial}
\newcommand{\pt}{\tilde p}

\newcommand{\ee}{\end{eqnarray}}


\newcommand {\be}[1]{
      \begin{eqnarray} \mbox{$\label{#1}$}  }

The concept of fractional exclusion statistics (FES) proposed by Haldane 
in 1991
\cite{haldane1}, has proved to be a useful concept to describe the 
statistical mechanics
and thermodynamics of certain  one-dimensional models. 
\cite{FESrealiz,wu1,veigy1,johnson1}

The basic idea of FES is that adding a number of particles, $
\Delta N$, to a system, blocks
$\Delta d$ of the states available  for the next particle according to the 
linear relation
$\Delta d = -g\Delta N$. Intuitively this corresponds to a 
repulsion between the particles,
but only very special types of interactions give rise to this type 
of exclusion of single
particle states. In fact all established examples of FES are in one dimension 
(or are effectively
one-dimensional like charged particles  restricted to the lowest Landau 
level by a strong
magnetic field \cite{veigy1,johnson1,hansson1}.) 
Thus, the observation\cite{bhaduri1,bhaduri2}
that, in a Thomas-Fermi approximation, a 
two dimensional Fermi or Bose gas with short range repulsive 
interactions has the same energy and number
density as an ideal FES gas 
(treated in the same approximation), deserves further 
study. 
It is not obvious why this kind of
interaction should give rise to exclusion of states in the sense of 
Haldane, and it is the
purpose of this note to provide a statistical mechanics derivation.

We start from the two-dimensional Hamiltonian
\be{ham}
H = \sum_{i=1}^N\left(\frac {p_i^2} {2m} + V(\vec r_i) \right) 
  +\frac {2\pi\hbar^2} m g \sum_{i<j}^N\delta^2(\vec r_i - \vec r_j )
\ee
that has been used as a model for atoms in Bose condensation experiments 
using highly
assymetrical traps
\cite{haugset1,assym}. The particular form of the delta function is chosen 
as to reproduce the
s-wave scattering phase-shifts in three dimensions, and the dimensionless 
coupling $g$ is
simply related to the corresponding scattering length and the out of 
plane extension of the
assymetrical trap \cite{haugset1,bhaduri2}. 
We shall assume that the temperature
is sufficiently high above the transition temperature that the only 
relevant mean field is the density, $n$, and that the fluctuations 
are small enough to be ignored. We shall return to these issues at 
the end of the paper.

Before we analyze the statistical mechanics of \pref{ham}, we shall 
give a simple thermodynamic argument for why, at the mean field level, 
we expect exclusion statistics. For simplicity we consider the case
with a constant external potential $V$, so that the
density, $n$, is also  constant. 
In a mean field approximation, and for a fixed number of particles, 
the interaction energy term in \pref{ham} just amounts to a constant 
shift of the energy density,
\be{eos}
{\cal E} &=& {\cal E}_{Free Bos.} +  \half kT \lambda_T^2 n^2 g  \\
&=&\frac {kT} {\lambda_T^2} \sum_{p=1}^\infty \frac 
{ B_{p-1} }{p!}  (\lambda_{T}^{2}n)^{p}
 + \half kT \lambda_T^2 n^2 g\, , \nonumber
\ee
where $kT=1/\beta$ is the temperature, and $\lambda_T 
= \sqrt{2\pi\hbar^2\beta/m}$ 
is the thermal wavelength. 
We  have used that for a free Bose gas in two
dimensions the pressure equals the energy density and 
substituted the pertinent virial expansion as expressed in the Bernoulli
numbers $B_n$\cite{sen1}. This expression is consistent with the system
being an ideal FES gas in two dimensions which is known to have 
a pressure equal to the energy density, and which differs from a 
free Bose gas only by a shift $\half g \lambda_T^2$ in the second 
virial coefficient\cite{isakov4}.

Let us now consider the statistical mechanics 
of \pref{ham}, and assume that the potential $V$ is
slowly varying compared with the thermal wavelength, $\lambda_{T}$. 
We  then divide
the system into cells of area $a^2$, where 
$\lambda_T \ll a \ll |\vec\nabla V /V|$, and study the 
statistical mechanics in each cell.     
In a mean field approximation,  the one-body Hamiltonian 
in the cell $\ell$  becomes,
\be{oneh}
H_{\ell} = \frac {p^2} {2m} + V(\vec r_\ell) 
       + \frac {2\pi\hbar^2} m g\, n(\vec r_\ell) 
\ee
where we  approximate the potential in the box with the constant 
$V(\vec r_{\ell})$ with $\vec r_\ell$  the position of the center of the 
cell. 
Also $n(\vec r_{\ell})=N_{\ell}/a^{2}$ is the mean number density in the cell
 $\ell$, with $N_{\ell}\gg 1$  the corresponding average number of particles.
 
The total energy, $E_{\ell}$ is, as usual, not simply the sum of the 
one particle energies $\epsilon_{i}^{\ell}$ of \pref{oneh}, but given by   
\be{toten}
E_{\ell} = \sum_{i}\left[\epsilon_{i}^{\ell}
- \frac {\pi\hbar^2} m g\, n(\vec r_\ell)\right] \, ,  
\ee  where the last term 
compensates for the double counting of the interaction energy. 

The number of available one particle quantum
states in the box, $d_\ell$, below some energy, $\epsilon^{\ell}$, given 
that there are already $N_{\ell}$ particles present, follows from \pref{oneh}
\be{nstat}
d_\ell = \frac {ma^2} {2\pi\hbar^2}{\epsilon}^{\ell}_{kin} = 
\frac {ma^2} {2\pi\hbar^2}  [\epsilon^{\ell} - V(r_\ell)] - gN_\ell \, ,
\ee 
with $\epsilon^{\ell}_{kin}$ the kinetic energy.
Note that $g=1$ corresponds to free fermions, and for general $g$, 
this relation immediately hints at exclusion statistics; the 
number of states in the box
decreases linearly with $N_\ell$. For a Hamiltonian of 
the form \pref{oneh}, this is true only in two dimensions. 

We can, however, still not conclude that our system is identical to an 
ideal FES gas.
Haldane's original definition of FES was for systems with a finite 
dimensional single
particle Hilbert space, but it was later generalized as to include 
ideal gases, and the
corresponding distribution functions were derived \cite{isakov1,wu1}. 
We shall now
demonstrate that the box Hamiltonian $H_\ell$ in \pref{oneh} indeed 
describes an ideal FES gas. 

Because of \pref{toten}, 
a microstate in the box can be labelled by a set of integers 
$0\le k_1 \le k_2 \cdots
\le k_{N_\ell}$, with the corresponding energy,
\be{micen} 
E_\ell =\sum_{i=1}^{N_\ell} \left( V(\vec r_\ell) 
       + \frac {2\pi\hbar^2} {ma^2} [k_i + \frac g 2 N_\ell ]  
       \right) \, .
\ee
Note that the integers $k_i$ are {\em not} proportional to the momenta; 
they are proportional  to the kinetic energy and  
correspond to the number of states 
below the  energy $\epsilon^{\ell}_{i}$. Of course, the real energy 
eigenvalues are not equally spaced, but we shall assume that the box 
is large enough for this effect to be negligible.

Next introduce the quantities $\tilde k_i$ by
\be{psenlab}
\tilde k_i = k_i + g\sum_j \theta(\tilde k_j - \tilde k_i)
\ee
with $\theta(x)$ being the step function,
and rewrite the energy as a sum of "pseudo energies", 
$\tilde\epsilon_{i}^{\ell}$, 
\be{psen}
E_\ell &=& \sum_{i=1}^{N_\ell} \tilde\epsilon^\ell_i \\
\tilde\epsilon^\ell_i &=& V(r_\ell) + \frac {2\pi\hbar^2} {ma^2} \tilde 
k_i \, .  \nonumber
\ee
The exclusion properties of this system are now manifest since \pref{psenlab} 
implies that
the pseudo energies must satisfy,
\be{excl}
\epsilon^{\ell}_{i+1}\ge \epsilon^{\ell}_i + \frac{2\pi}{ma^2} \hbar^2 g.
\ee 

It should be noted that the relation \pref{nstat} holds  only  because 
the kinetic energy and the number density scale as the same power of 
the box size $a$. 
This is true for the present case of particles with quadratic 
dispersion in two dimensions, 
but also for particles in one dimension with linear dispersion, which
is the case for anyons in the
lowest Landau level, or equivalently, chiral particles on a circle with an 
$N^2$ type interaction \cite{hansson1}. In fact, the two models studied in 
this paper and in \cite{hansson1} can be exactly mapped onto each 
other by identifying the $\tilde k_{i}$ in \pref{psenlab} with the 
``pseudomomenta'' introduced in \cite{hansson1}.

The interaction strength, $g$ only enters in the combination
$\alpha = g\hbar^{2}$. Recently it was shown that by taking the
limit $g\rightarrow \infty$, $\hbar \rightarrow 0$ with $\alpha$ fixed, 
$\alpha$ can be interpreted as a {\em classical} exclusion statistics 
parameter\cite{hansson2000}. 
Using the  Thomas-Fermi approximation for the system \pref{ham}, 
and taking the  high $T$ limit, it is easy to show that 
the density is given by $n(\vec r) = n_B(\vec r)
\left(1 + \alpha \frac {2\pi} {kT} n_B(\vec r)\right)^{-1}$, where 
$n_B(\vec r)$ is the density of a non-interacting Stefan-Boltzmann gas 
in the same potential  $V(\vec r) $.
Note that all $\hbar$-dependence is gone, and that the classical 
density is lowered because of a {\em classical} statistics effect. 

The formal proof that the exclusion property \pref{excl} corresponds 
to an ideal FES gas as
defined in \cite{isakov1} is a straightforward modification of the
one given in \cite{isakov2} for a multispecies system in the
fermionic representation: Going to a continuum description with
``momenta'' $p_i$ and ``pseudomomenta'' $\pt_i$ defined as
\be{ppt}
p_i = \left( \frac{2\pi\hbar}{a} \right)^2 k_i ; \ \ \ \ \
\pt_i = \left( \frac{2\pi\hbar}{a} \right)^2 {\tilde k}_i,
\ee
replacing the sums over $k$ and $\tilde k$ by integrals in the 
usual way and denoting the corresponding particle densities
in momentum space by $\nu(p)$ and $\rho({\tilde p})$, respectively,
one finds the continuum version of Eq.\pref{psenlab},
\be{psenc}
\pt = p + g \int d\pt' \rho(\pt') \theta(\pt - \pt') \ .
\ee
Furthermore one has to demand conservation of the number of particles
when changing variables from $p$ to $\pt$, \ie 
$\nu(p) dp = \rho(\pt) d\pt$. Combining this with Eq.\pref{psenc} gives
\be{nurho}
\nu(p) = \frac{\rho(\pt)}{1 - g\rho(\pt)}.
\ee
Inserting this into the standard expression for the bosonic
non-equilibrium entropy,
\be{ent1}
S = -k\left( \frac{a}{2\pi\hbar} \right)^2
      \int dp \left[ \nu\ln\nu - (1+\nu)\ln(1+\nu) \right]
\ee
exactly reproduces the entropy of a ideal FES gas \cite{isakov1,wu1},
\be{ent2}
S &=& -k \left( \frac{a}{2\pi\hbar} \right)^2
         \int d\pt \left[ \rho\ln\rho \right.    \\
      &-& \left. \left( 1-(g-1)\rho \right) \ln \left( 1-(g-1)\rho \right)
         \right. \nonumber \\
      &+&\left. \left( 1-g\rho \right)\ln\left( 1-g\rho \right) 
      \right] \nonumber
\ee
from which all thermodynamics follows.

Since we have explicitly ignored both the possibility of a quantum 
condensate and of pairing fields other than the density, 
the results of this paper (and those of Ref. 
\cite{bhaduri2}) can not be used for temperatures below or in the 
vicinity of the Bose condensation transition $T_{c}$. This is true
irrespective of whether this transition is of the 
Kosterlitz-Thouless 
type or not\cite{petrov2000}. 
Rather our results should be relevant in a temperature 
regime where the exclusion statistics, due to the repulsive interaction,
corresponds to a small correction to the ideal Bose gas.
It is an interesting 
open question whether the quasi particles above a 
two-dimensional bose condensate also can be described using 
exclusion statistics. To answer this question one would analyze the 
corresponding statistical mechanics in a more sophisticated mean 
field approximation that includes effects of phase coherence and 
pairing mean fields\cite{griffin}.

We should finally  point out that while our analysis was entirely in the 
context of mean field approximations, it is an interesting question 
whether or not the  full quantum problem  
of a two dimensional gas with delta function\cite{2d-delta} 
also allows a description 
in terms of exclusion statistics in some range of temperatures.  
Since the interaction does not involve 
any dimensional parameter,  this possibility is not excluded.

\vskip 2mm
\noi {\bf Acknowledgement}: We thank St\'ephane Ouvry,  Alexis 
Polychronakos and Gora Shlyapnikov for useful discussions.

\vspace{-0.5cm}
\bibliographystyle{unsrt}

\vspace{0.5cm}
\end{multicols}
\end{document}